\documentclass[amsmath,amssymb,superscriptaddress]{revtex4-2}
\usepackage{amssymb}
\usepackage{amsfonts}
\usepackage{amsmath}
\def\be{\begin{equation}}
\def\ee{\end{equation}}
\def\arr{\begin{array}{rll}}
\def\ea{\end{array}}
\def\bea{\begin{eqnarray}}
\def\eea{\end{eqnarray}}
\begin{document}
\title{Integrable isotropic profiles for polarized light }
\author{Mher Davtyan}
\email{mher.davtyan@gmail.com}
\affiliation{Institute of Radiophysics and Electronics, Ashtarak-2, 0203, Armenia   }
\author{Zhyrair Gevorkian}
\email{gevork@yerphi.am}
\affiliation{Yerevan Physics Institute, 2 Alikhanian Brothers St., Yerevan  0036 Armenia}
\affiliation{Institute of Radiophysics and Electronics, Ashtarak-2, 0203, Armenia   }
\author{Armen Nersessian}
\email{arnerses@yerphi.am}
\affiliation{Yerevan Physics Institute, 2 Alikhanian Brothers St., Yerevan  0036 Armenia}
\affiliation{Bogoliubov Laboratory of Theoretical Physics, Joint Institute for Nuclear Research, Dubna, Russia}
\affiliation{Institute of Radiophysics and Electronics, Ashtarak-2, 0203, Armenia   }

\begin{abstract}
We consider the propagation of polarized light in the medium with  isotropic refraction index profile and  show that polarization violates the additional symmetries of the medium. Then we suggest a scheme for the construction of   polarization-dependent  refraction index which restores all symmetries of the initial profile. We illustrate the proposed scheme on the examples of Luneburg and Maxwell's fisheye profiles.
\end{abstract}

\maketitle

\section{Introduction}
It is well-known that the minimal action principle came in physics from geometric optics. Initially, it was invented for the description of the propagation of light and is presently known as the Fermat principle
\begin{equation}
 {\cal S}_{Fermat} =\frac{1}{\lambda_{0}}\int n(\mathbf{ r}) \vert\frac{d\mathbf{r}}{d\tau} \vert d\tau
\label{gactions2}
\end{equation}
where  $n(\mathbf{ r})$ is the
refraction index, and $\lambda_0$ is the wavelength in vacuum. This  action
could be interpreted as the action of the system on the
three-dimensional curved space equipped with the ``optical metrics"  of Euclidean signature (see. \cite{arnold})
 \begin{equation}
  d{\tilde l}^2= n^2(\mathbf{r})d\mathbf{r} \cdot d\mathbf{r}\;.
\label{om}
\end{equation}
Thus, the symmetries of the system which describe the propagation of light in a particular medium are coming from the symmetries of the respective optical metrics.
On the other hand,  in accordance with Maupertuis  principle, one can relate any non-relativistic systems describing with the Lagrangian
\be
\mathcal{L}_0=\frac12 g(\mathbf{r})\dot{\mathbf{r}}^2 - V(\mathbf{r}),
\label{nonrel}\ee
can be related with the action \eqref{gactions2}, with the refraction index ( $E$ is the value of the system's energy)
\be
 n({\bf r})=\lambda_0\sqrt{2g({\bf r})(E-V({\bf r}))}.
\label{rih} \ee
Clearly such an optical system inherits  all symmetries of the initial system.

 In the superintegrable systems, i.e. in the systems with a maximal number of functionally independent integrals   ($2N-1$ integrals for $N$-dimensional system), all the trajectories   become closed. The closeness of the trajectories makes respective optical  profiles highly relevant in the study of cloaking and perfect imaging phenomena.

The most well-known profile of this sort is the so-called ``Maxwell's fisheye" profile which is defined by the metrics of (three-dimensional) sphere or pseudosphere (under pseudosphere we mean the upper (or lower) sheet of the two-sheet hyperboloid).
\begin{equation}
n_{Mfe}(\mathbf{r}) =\frac{n_0}{|1 +  \kappa  \mathbf{r}^2 |}, \qquad \kappa=\pm\frac{1}{4r^2_0}
\label{Mfe0}
\end{equation}
where the plus/minus sign in the expression for $\kappa$ corresponds to the sphere/pseudosphere with the radius  $r_0$, and $n_0>0$.
Apart from applications in cloaking and perfect imaging phenomena \cite{Pendry06}, Maxwell's fish eye has numerous different applications, see, e.g.
 \cite{perczel2018}

Another well-known example of a superintegrable profile  is the Luneburg lens, which is related with three-dimensional isotropic oscillator,
\begin{equation}
n_{Lun}(\mathbf{r}) = n_0 \sqrt{1 - \left(\frac{\mathbf{r}}{r_0}\right)^2}.
\label{lun}
\end{equation}
However, the optical systems describing by the action \eqref{gactions2} do not take into account polarization of light.

 Introduction of spin (polarization) results to the rotation of this plane by a constant angle proportional to spin, moreover,  it breaks the non-rotational symmetries of the optical systems superintegrable
 integrable profile, so that photon trajectories no longer remain closed \cite{gdn}.
 In these systems, the ray trajectories belong to the plane which is orthogonal to the angular momentum.  Thus, the key property  of superintegrable isotropic profiles  which makes them  relevant in cloaking and perfect imaging phenomena becomes violated.

In this paper, following \cite{mfePolarized} we propose  a general scheme of the deformation of  isotropic refraction index profiles. It allows us to restore the initial symmetries of the system after one takes the light polarization into consideration. Namely,
to  preserve the qualitative properties of scalar wave trajectories for the propagating polarized light, we suggest replace the   refraction index $n(r)$
with the   modified index $n^s(r)$ which is the solution
(with respect to $p$) of the following equation:
\be
p=\frac{1}{\lambda_0} n\left(\sqrt{r^2+\frac{s^2}{p^2}}\right),\quad\Rightarrow\quad p=\frac{1}{\lambda_0}n^s(r),
\label{gs}\ee
where $s$ is polarization of light.

For the particular cases of Maxwell's fisheye and Luneburg profiles it yields the following expressions
\be
n^{s}_{Mfe}(\mathbf{r})= \frac{n_{Mfe}(\mathbf{r})}{2} \left(1 + \sqrt{1 - \frac{( 2s\lambda_0)^2 }{n_0}\frac{\kappa}{n_{Mfe}(\mathbf{r})}}\right),
\label{spinFishEye0}
\ee
\be
n^{s}_{Lun}(\mathbf{r})= \frac{n_{Lun}(\mathbf{r})}{\sqrt{2}}\sqrt{ 1 + \sqrt{1 + \left(\frac{2s \lambda_0 n_0}{r_0n^{2}_{Lun}(\mathbf{r})}\right)^2} }.
\label{spinLuneburg}
\ee
where $n_{Mfe}(\mathbf{r})$  and $n_{Lun}(\mathbf{r})$ are  the original Maxwell's fisheye  and Luneburg profiles given, respectively, by  by \eqref{Mfe0}and \eqref{lun}, while $s$ is the light polarization.
The proposed deformations restores all the symmetries of the initial systems  with Maxwell's fisheye  and Luneburg profiles, which were broken after the inclusion of polarization.

Let us notice, that for the inclusion of polarization one should   extend the initial physical space by additional, isospin degrees of freedom  \cite{horvathy} (see also \cite{deriglazov} and refs therein), which seems intuitively artificial. Otherwise, one can extend the initial optical Lagrangian by the additional terms depending on higher derivatives \cite{aghamalyan}. However, the latter, aesthetically attractive, approach describes very particular class of optical systems.
By this reason in our study we will mostly use    the Hamiltonian framework, where taking into account of polarization is very natural.

The paper is organized as follows.
In {\sl Section 2}, we present the  Hamiltonian formulation of the  optical system given by the action \eqref{gactions2}.
In {\sl Section 3}, we present the Hamiltonian formalism for the polarized light propagating in an optical medium and propose the general scheme of the deformation of isotropic refraction index   which allows us to restore the initial symmetries after the inclusion of polarization.
%
%
%

Through the text we will use the notation $r:=|\mathbf{r}| $, $\mathbf{r}:=(x_1,x_2,x_3)$, $\mathbf{p}:=(p_1,p_2,p_3)$, $p:= |\mathbf{p}| $, and so on.

 \section{Hamiltonian formalism for geometric optics}

Due to reparametrization-invariance of the action   \eqref{gactions2},  the Lagrangian is singular
 and the constraint between  momenta  and coordinates appears there

\begin{equation}
\Phi:= \frac{{\bf p}^2}{n^2({\bf r})} -\lambda^{-2}_0 =0.
\label{constraint}
\end{equation}
Hence, in accordance with the Dirac's constraint theory 
the respective  Hamiltonian system
is defined by the
canonical Poisson brackets
\begin{equation}
\{x_i, p_j\}=\delta_{ij},\quad   \{p_i, p_j\}=\{x_i, x_j\} =0,
\label{pb0}\end{equation}
and by the  Hamiltonian
\begin{equation}
\mathcal{H}_0=\alpha({\bf p},{\bf r})\Phi =\alpha({\bf p},{\bf r})\left( \frac{{  p}^2}{n^2({\bf r})}-\lambda^{-2}_0 \right)\approx 0.
\label{h0}
\end{equation}
Here $\alpha$ is the Lagrangian  multiplier  which could be an arbitrary function of coordinates and momenta, and $i,j=1,2,3$.
The notation ``weak zero", $\mathcal{H}_{0}\approx 0$,  means that when writing down the Hamiltonian equations of motion,
we should take into account the constraint \eqref{constraint} only after the differentiation.
The  arbitrariness in the choice of the function $\alpha$ reflects the reparametrization-invariance of  \eqref{gactions2}.
For the description of the  equations of motion in terms of arc-length of the original Euclidian space one should choose (see,  e.g.  \cite{bliokh})
\begin{equation}
\alpha =\frac{n^2({\bf r})}{ {  p} + \lambda^{-1}_0 n({\bf r})},\qquad \Rightarrow\quad \mathcal{H}_{\rm Opt}= p - \lambda^{-1}_0n({\bf r}).
\label{alpha}
\end{equation}
With this choice, the  equations of motion take the conventional form \cite{ko}
 \begin{equation}
\frac{d{\bf {p}}}{dl}=\lambda^{-1}_0{\bf \nabla} n({\bf r}),\qquad
\frac{d{\bf {r}}}{dl}=\frac{\bf p}{ {  p} },
\label{hameq}
\end{equation}
where $dl:=\alpha({\bf r}, {\bf p})d\tau$ is the element of arc-length.
These equations describe the motion of a wave package with center coordinate ${\bf r}$ and momentum ${\bf p}$ in the medium with refraction index
$n({\bf r})$.\\

The Hamiltonian formulation of the system \eqref{nonrel}  is given by canonical    Poisson bracket \eqref{pb0} and by the Hamiltonian
\be
H=\frac{{  p}^2}{g({\bf r})}+V({\bf r}).
\label{h}\ee
In accordance with the Mopertuit principle,  fixing the energy surface $H=E$,  we can express the momentum $p$ via coordinates $\mathbf{r}$
  relate its trajectories with the optical Hamiltonian \eqref{h0} defined by the refraction index \eqref{rih}
Clearly, the optical Hamiltonian \eqref{h0} (as well as the Hamiltonian  \eqref{alpha}) with the  refraction index  \eqref{rih} inherits all symmetries and constants of motion of the Hamiltonian \eqref{h}. Canonical transformations preserve the symmetries of the Hamiltonians and their level surfaces. Hence, we are able to construct the physically non-equivalent optical Hamiltonians (and refraction indices) with the same symmetry algebra.

\section{Polarized light}
When taking into account light polarization we should add to the  scalar Lagrangian $L_0={\bf p\dot{r}}-p+\lambda_0^{-1}n$ the additional term $L_1=-s\mathbf{A}( \mathbf{p})\dot{\mathbf{p}}$,
where $s$ is spin of photon, and   $\mathbf{ A}$ is the
the vector-potential  of ``Berry monopole"  (i.e. the
  potential of the magnetic (Dirac) monopole located at the origin of momentum space) \cite{bliokh}
  \be
\mathbf{F}:=\frac{\partial}{\partial \mathbf{ p}}\times \mathbf{A}(\mathbf{p})=\frac{\mathbf{p}}{  {p} ^3}
\label{F}\ee

From the  Hamiltonian viewpoint this means to preserve the form of the Hamiltonian \eqref{h0}
and  replace the  canonical Poisson brackets \eqref{pb0}
by the twisted  ones
 \begin{equation}
  \{x_i, p_j\}=\delta_{ij},\qquad \{x_i, x_j\}=s\varepsilon_{ijk}F_k(\mathbf{p}), \qquad\{x_i, x_j\} =0,
 \label{pbB}\end{equation}
  where $ i,j,k=1,2,3$, and  $F_k$ are  the components of Berry monopole \eqref{F}.
On this phase space the rotation generators take  the form
  \be
\mathbf{J}=\mathbf{r}\times \mathbf{p}+ s\frac{{\mathbf p}}{ {p} }
\label{Js}
\ee
while
the equations of motion read
  \be
\frac{d\mathbf{p}}{dl}=\lambda_0^{-1}\mathbf{\nabla} n(\mathbf{r}),\qquad \frac{d\mathbf{r}}{dl}=\frac{\mathbf{p}}{ p } - \frac{s}{\lambda_0}\mathbf{F} \times \mathbf{\nabla}n({\bf r}),
  \ee
However, the above procedure, i.e. twisting the Poisson bracket with preservation of the Hamiltonian,  violates  the non-kinematical (hidden) symmetry of the system.
To get the profiles admitting the symmetries in the presence of polarization, we use the following observation \cite{lnp}
 . Assume we have the three-dimensional rotationally-invariant system
 \be
 \mathcal{H}_0=\frac{{p}^2}{2g(r)}+V(r),\qquad  \{p_i, x_j\}=\delta_{ij}, \quad  \{p_i, p_j\}=\{x_i, x_j\}=0.
 \ee
For the inclusion of  interaction with magnetic monopole, we should  switch from the canonical Poisson brackets to the twisted ones:
\be
\{p_i, x_j\}=\delta_{ij}, \qquad  \{p_i, p_j\}=s\varepsilon_{ijk}\frac{x_k}{r^3},\qquad \{x_i, x_j\}=0.
\label{PBsp}\ee
The rotation generators  then read
\be
\mathbf{J}=\mathbf{r}\times \mathbf{p}+ s\frac{{\mathbf r}}{r}\;: \quad \{J_i,J_j\}=\varepsilon_{ijk}J_k.
\label{Jr}\ee
By modifying the initial Hamiltonian to
\be
\mathcal{H}_s=\frac{{p}^2}{2g( r)}+\frac{s^2}{2g(r){r}^2}+V(r),
\label{Hs}\ee
we find that trajectories of the system preserve their form, but the plane which they belong to, fails to be orthogonal to the
the axis $  {\mathbf{J}} $. Instead, it  turns to the constant angle
\be
\cos\theta_0=\frac{s}{|\mathbf J|}.
\ee
For the systems with hidden symmetries one can find the appropriate modifications of the hidden symmetry generators respecting the inclusion of the monopole field.

For applying this observation on the systems with polarized light,   we should choose the appropriate integrable system with magnetic monopole, and then perform
simple canonical transformation which yields the Poisson brackets for polarized light \eqref{PBsp} :
\be
({\bf p}, {\bf r})\to (- {\bf r} , {\bf p}).
 \label{ct}\ee

 Afterwards we need to solve the following equation
 \be
  r^2+\frac{s^2}{p^2}-2 g(p) (E-V(p))=0,\quad\Rightarrow\quad p=\frac{n^s_{inv} ( r)}{\lambda_0}.
 \ee
So, to  preserve the qualitative properties of scalar wave trajectories for the propagating polarized light, we should replace it with the modified index $n^s(r)$ which is the solution
(with respect to $p$) of the following equation:
\be
p=\frac{1}{\lambda_0} n\left(\sqrt{r^2+\frac{s^2}{p^2}}\right),\quad\Rightarrow\quad p=\frac{n^s(r)}{\lambda_0},
\label{gs}\ee
where $s$ is polarization of light.

For example, to get the ``polarized Coulomb profile"  we have  to start  from the free-particle Hamiltonian on three-dimensional  sphere/hiperboloid
Then, after fixing the energy surface $H_{s}=E$ and performing canonical transformation \eqref{ct}
we arrive to the third-order algebraic equation which
  has either one real and two complex solutions or three real solutions, which  describe the ``polarized Coulomb profiles".
Conversely, when we start from the Coulomb problem  we will arrive to the ``polarized Maxwell's fish eye" \eqref{spinFishEye0}, i.e. the deformation of the ``Maxwell fish eye"  which preserves,  in the presence of polarized light, all symmetries of initial scalar system.
While to get "polarized Luneburg profile" \eqref{spinLuneburg} we should choose, to the role of initial Hamiltonian, the isotropic oscillator.

\vspace{5mm}

{\sl Acknowledgements.}
M.D. thanks the organizers of XVIII Conference "Symmetry Methods in Physics" (10-17.07.2022  Yerevan) for given opportunity to present this work.  
Authors acknowledge partial  financial support from
Armenian Committee of Science, projects 20RF-023, 21AG-1C062 (A.N., Zh.G.)  and  from the Russian Foundation
of Basic Research   grant 20-52-12003 (A.N.).

\end{document}